\documentclass[a4paper,11pt]{article}

\usepackage{jinstpub} 

\usepackage{wasysym}
\usepackage{xspace}
\usepackage{xcolor}
\usepackage{color}
\usepackage[normalem]{ulem}

\usepackage{lineno}

\newcommand{\dvthgem}{\ensuremath{\Delta V_{\textrm{THGEM}}}\xspace}

\newcommand{\eextraction}{\ensuremath{E_{\textrm{extraction}}}\xspace}
\newcommand{\edrift}{\ensuremath{E_{\textrm{drift}}}\xspace}

\newcommand{\sone}{S1\xspace}
\newcommand{\soneprime}{S1'\xspace}
\newcommand{\stwo}{S2\xspace}
\newcommand{\ETEpe}{ETE\ensuremath{_{\textrm{pe}}}\xspace}
\newcommand{\ETEie}{ETE\ensuremath{_{\textrm{ie}}}\xspace}
\newcommand{\Wsc}{\ensuremath{W_{sc}}\xspace}

\newcommand{\ntwo}{N\ensuremath{_2}\xspace}
\newcommand{\lntwo}{L\ntwo}
\newcommand{\am}[0]{$^{241}$Am\xspace}

\newcommand{\registered}{\ensuremath{^{\textrm{\textregistered}}}\xspace}

\newcommand{\ignoreblock}[1]{}

\newcommand{\original}[1]{}
\newcommand{\changes}[1]{#1}
\title{\boldmath Electron transfer efficiency in liquid xenon across THGEM holes}

\author[a,b,1]{G. Mart\'inez-Lema\note{Equal contributor},}
\author[a,b,1]{A. Roy,}
\author[b]{A. Breskin}
\author[a,2]{and L. Arazi \note{Corresponding author.},}

\affiliation[a]{Unit of Nuclear Engineering, Ben Gurion University of the Negev, Beer-Sheva, Israel}
\affiliation[b]{Department of Particle Physics and Astrophysics, Weizmann Institute of Science, Rehovot, Israel}

\emailAdd{larazi@post.bgu.ac.il}

\abstract{Dual-phase liquid-xenon time projection chambers (LXe TPCs) deploying a few tonnes of liquid are presently leading the search for WIMP dark matter. Scaling these detectors to 10-fold larger fiducial masses, while improving their sensitivity to low-mass WIMPs presents difficult challenges in detector design. Several groups are considering a departure from current schemes, towards either single-phase liquid-only TPCs, or dual-phase detectors where the electroluminescence region consists of patterned electrodes. Here, we discuss the possible use of Thick Gaseous Electron Multipliers (THGEMs) coated with a VUV photocathode and immersed in LXe as a building block in such designs. We focus on the transfer efficiencies of ionization electrons and photoelectrons emitted from the photocathode through the electrode holes and show experimentally that efficiencies approaching 100\% can be achieved with realistic voltage settings. The observed voltage dependence of the transfer efficiencies is consistent with electron transport simulations once diffusion and charging-up effects are included.

}


\arxivnumber{1234.56789} 

\begin{document}
\maketitle
\flushbottom

\section{Introduction}
\label{sec:intro}

Noble liquid time projection chambers (TPCs) are among the leading instruments in the field of rare event searches. In particular, the most successful dark matter (DM) experiments to date employ large volumes of liquid xenon (LXe) and liquid argon (LAr), in dual-phase TPCs instrumented with photosensors. The working principle of these apparatuses is based on the detection of two light signals emitted when a particle interacts in the active volume: a prompt scintillation light (\sone), proportional to the number of excited states created at the site of interaction, and a delayed electroluminescence (EL) signal (\stwo), proportional to the number of ionization electrons produced in the interaction, which is generated when the electrons are extracted into the gas phase under an intense electric field.

As previous experiments (up to $\sim$1 tonne of active mass) yielded null results \cite{PhysRevD.98.102006,PhysRevD.96.022008,PhysRevLett.123.251801,PhysRevD.102.072004,PhysRevD.98.062005,PhysRevLett.122.131301,YUAN2022137254}, larger and more sensitive detectors have been constructed and begun taking data \cite{PhysRevLett.129.161805,2022arXiv220703764A,PhysRevLett.129.161803,PhysRevLett.129.161804}.
For the next, and possibly final stage of direct DM searches utilizing this technology, the DARWIN collaboration aims to build a 50-tonne LXe detector to cover the accessible parameter space, down to the so-called irreducible neutrino floor \cite{Aalbers_2016}. Parallel efforts with liquid argon (LAr) aim at a 300-tonne DM detector (``ARGO'') \cite{argo}, initially passing through the 20-tonne DarkSide-20k \cite{Aalseth_2018}.

While dual-phase TPCs have been highly successful large-mass detectors, their upgrade towards the $>$50-tonne scale implies facing serious challenges. One particular concern is the requirement to maintain an undisturbed liquid-gas interface in between and parallel to the anode and gate meshes, which are typically positioned 5-10 mm apart. This is crucial for achieving a uniform \stwo response across the active volume, which translates directly to the energy resolution and signal/background discrimination capability of the detector. Meeting this requirement becomes increasingly challenging for large TPC diameters ($>2$~m) due to sagging induced by electrostatic forces between the meshes. A second concern is the need for unambiguous identification of low \sone signals, particularly low-energy nuclear recoils where the number of photoelectrons detected may be too small to distinguish them from the inherent dark noise of the photosensors. This problem becomes more pronounced in larger detectors, where the total photosensitive area scales with the diameter squared. In this respect, although silicon photomultipliers (SiPMs) could offer a major improvement in radiopurity compared to photomultiplier tubes (PMTs), their high dark-count rate at 180~K \cite{ANFIMOV2021165162} presents a challenge for their application in LXe DM detectors, as it may require setting an undesirably high threshold for coincident detection of S1 photons.
A third problem in dual-phase TPCs is the delayed emission of electrons from the liquid-gas interface, which hinders searches for very-low-energy events relying on ``\stwo-only analysis'' \cite{LUX_single_e_emission2020,Kopec_2021,XENON1T_single_e_emission2022}.


Various concepts for both single-phase and dual-phase detectors are currently being explored or proposed to mitigate the problems associated with scaling up dual-phase TPCs. One possibility is to use thin wires to generate EL inside the liquid \cite{Aprile_2014, Giboni_2014, Lin_2021, Juyal_2021_wires, Kuger_2022, Wei_2022}. Another possibility is to use perforated electrodes immersed inside the liquid (Liquid Hole-Multipliers, LHMs \cite{Breskin_2013}), where electroluminescence signals are formed either inside a vapor bubble trapped below the electrode \cite{Arazi:2015uja, Arazi:2016nvi, Erdal:2015kxa, Erdal:2016jhl, Erdal:2017yiu, Erdal:2018bjg, Erdal:2019dkk, Erdal:2019wfb}, or on thin strips and micro-structured electrodes in a liquid-only configuration \cite{Breskin_2022}. LHMs coated with a VUV-sensitive photocathode (e.g., cesium iodide, CsI), can be used to generate large EL signals in response to single VUV photons, by focusing the photoelectrons emitted from the photocathode either into the vapor bubble or onto the thin strips. The same electrodes should also serve to transfer the ionization electrons into the EL region. 
The Floating Hole Multiplier (FHM) was recently proposed as a potential solution for reducing interface instabilities in dual-phase TPCs \cite{Chepel_2023}.
It consists of a perforated electrode floating freely on the liquid surface, with EL formed within the holes and their vicinity. Coating the electrode with CsI on the liquid side would provide both S1 and S2 signals. In the concept discussed here (which was partly presented in recent conferences \cite{lidine_2022,mpgd_2022}) we propose to transfer the S2 electrons and S1-induced photoelectrons from a CsI-coated perforated electrode immersed in the noble liquid through the holes and extract them to the gas phase. EL signals would originate from the holes of a second perforated electrode located in the vapor phase.


A key requirement of LHMs is that they provide high transfer efficiencies for both S1-induced photoelectrons and S2 ionization electrons across their holes. The transfer of ionization electrons across Thick Gaseous Electron Multipliers (THGEMs) immersed in LAr was studied extensively by Bondar et al. \cite{BONDAR2019162431,BUZULUTSKOV201829,DarkSide:2020oas,BONDAR2013213}, who demonstrated an electron transfer efficiency close to 100\%. In this work we extend their research by applying it to THGEMs immersed in LXe, focusing on the use of CsI photocathodes for improved \sone detection, and exploring a broader operational regime that fits better the conditions of a Xe-based detector. Our experimental results are compared to simulations showing the effect of electron diffusion and charging up.

\section{Experimental setup}
\label{sec:setup}

The measurements reported in this work were performed using the Mini Xenon cryostat (MiniX), a compact LXe system described in detail in \cite{Erdal:2015kxa}. The chamber is a cylindrical volume, 100~mm in diameter and 100~mm in height, filled with roughly 0.5~L of LXe. An outer volume kept in vacuum provides thermal insulation. The inner volume hosts the detector assembly, suspended from a flange on top, which also holds the high voltage feedthroughs to bias the system. A fused-silica viewport on the side, at an angle of $60^{\circ}$ with respect to the vertical axis allows a visual inspection of the setup and illumination by a VUV lamp. Part of the cryostat wall is made of a \lntwo-cooled, temperature-modulated copper structure, that allows for a controlled liquefaction of the gas. A recirculation system ensures continuous purification of xenon.
LXe is extracted through a tube at the bottom of the cryostat, fed into a heat exchanger, and circulated through a SAES hot getter, model PS3-MT3-R-2. After purification, the gaseous xenon returns to the top part of the cryostat through the heat exchanger.

The detector assemblies were mounted in a \ntwo environment to avoid degradation of the CsI photocathode, which was vacuum-deposited on the THGEM electrode in a dedicated evaporator. The inner volume of the cryostat was flushed with \ntwo while not in operation, to prevent moisture contamination.
The chamber was pumped down to $\sim10^{-6}$ mbar before filling with gaseous Xe and the results of the experiments reported here were obtained after several days of recirculation. The LXe temperature was kept constant at 175~K ($\pm 0.1$ K)  during the measurements.
\changes{The electron lifetime in the setup was not measured directly, but it was estimated to be considerably longer than the drift time.
This estimate was obtained by analyzing data from other experiments in the same setup where we did not observe changes in the amplitude of light and charge pulses over time already 1-2 days after starting Xe recirculation.
}

\begin{figure}
    \centering
    \includegraphics[height=55mm]{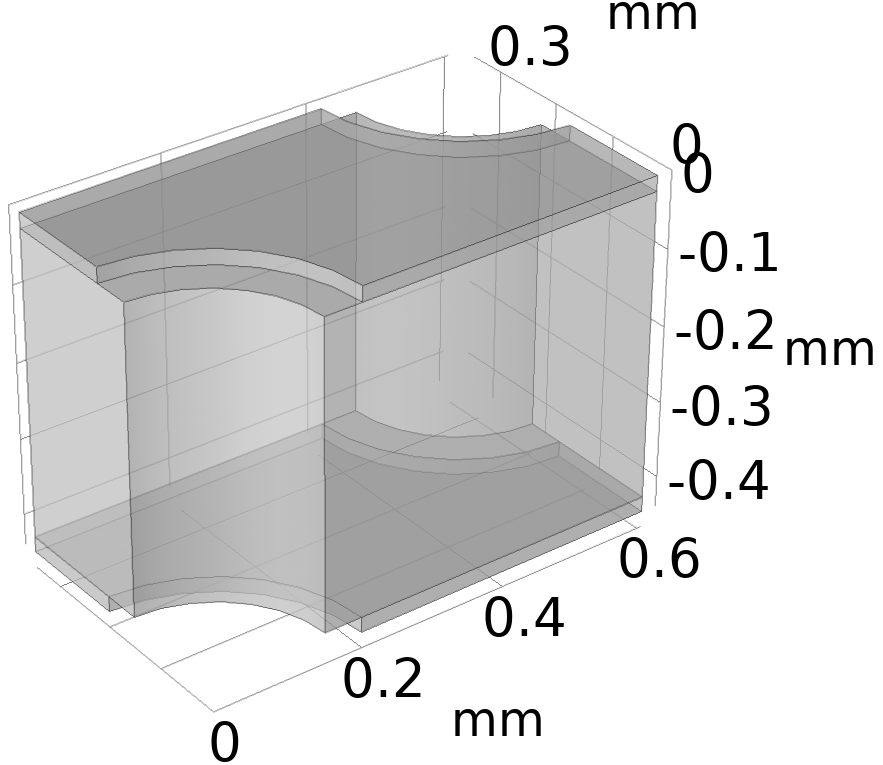}
    \includegraphics[height=55mm]{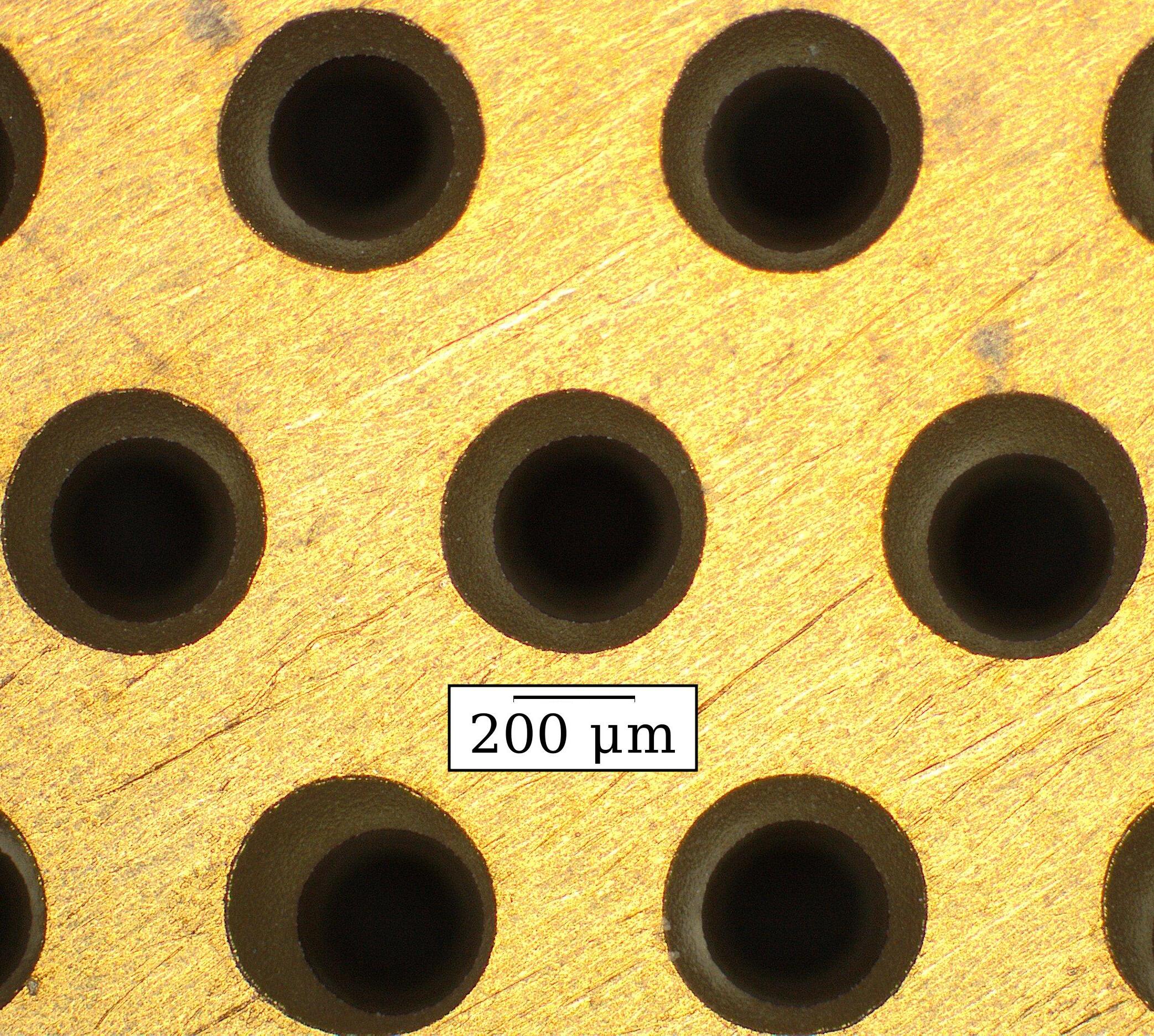}
    \caption{Left: 3D model of a unit cell of the THGEM used in the measurements reported in this article, built in COMSOL Multiphysics\registered.
    Right: Detail of the hole pattern of a THGEM under a microscope.
    }
    \label{fig:thgem}
\end{figure}

Each measurement required a slightly different configuration described below. All setups are comprised of a 33~mm diameter, 0.4~mm thick, gold-plated THGEM electrode made of FR4, with 0.3~mm diameter holes in a hexagonal pattern and a pitch of 0.7~mm. The holes in the insulator were surrounded by 50~$\mu$m rims. A 3D model of the THGEM unit cell built in COMSOL Multiphysics\registered and a picture of the hole pattern of a THGEM are shown in Figure \ref{fig:thgem}.
In addition to the THGEM, we used stainless-steel woven-mesh electrodes, with 50~$\mu$m wires forming a square pattern with a pitch of 500~$\mu$m. Every setup was mounted on PEEK rods to ensure insulation from the cryostat body.
The gaps between the different components were created using either PEEK or FR4 spacers. Currents were measured using a Keithley 610C solid-state analog electrometer.

\section{Measurement of the electron transfer efficiency}
\label{sec:ete}

\subsection{Methodology}
\label{sec:methodology}

The electron transfer efficiency (ETE) is defined as the ratio of the current read across the target electrode ($I_{target}$) to the total current measured from the source electrode ($I_{source}$), as shown in Eq.~\ref{eq:efficiency}.

\begin{equation}
    \textrm{ETE} = \frac{I_{target}}{I_{source}}.
    \label{eq:efficiency}
\end{equation}

\noindent
While the target electrode is the same in both cases (photoelectrons and ionization electrons), the source electrode is defined differently in each case.
The currents on each electrode were measured independently \changes{and consecutively}. To accommodate the requirements of the analog electrometer used for the measurements, which necessitates a grounded input electrode, the voltage ladder was adjusted for each measurement while maintaining consistent voltage differences.
Furthermore, since the charge multiplication threshold in LXe is approximately $\gtrsim$700~kV/cm \cite{Aprile_2014}, and the maximum electric field in our geometry is roughly one-tenth of that value, this ratio represents the likelihood of an electron reaching the target electrode.
\changes{
The currents produced by the alpha particle interactions and by photoelectron emission from CsI were of the order of $\sim$~100~pA.
The uncertainty of the measurement was determined by small fluctuations of the order of $\sim$~2~pA.
}

Note that due to the intense dipole field produced near the THGEM holes, which affects the field above and below the electrode, the electric fields reported here are not the nominal ``parallel-plate''-like values, but those obtained using a COMSOL simulation of the same geometry.

\subsection{Photoelectron transfer efficiency}
\label{sec:ete_s1prime}

\begin{figure}
    \centering
    \includegraphics[width=0.7\columnwidth]{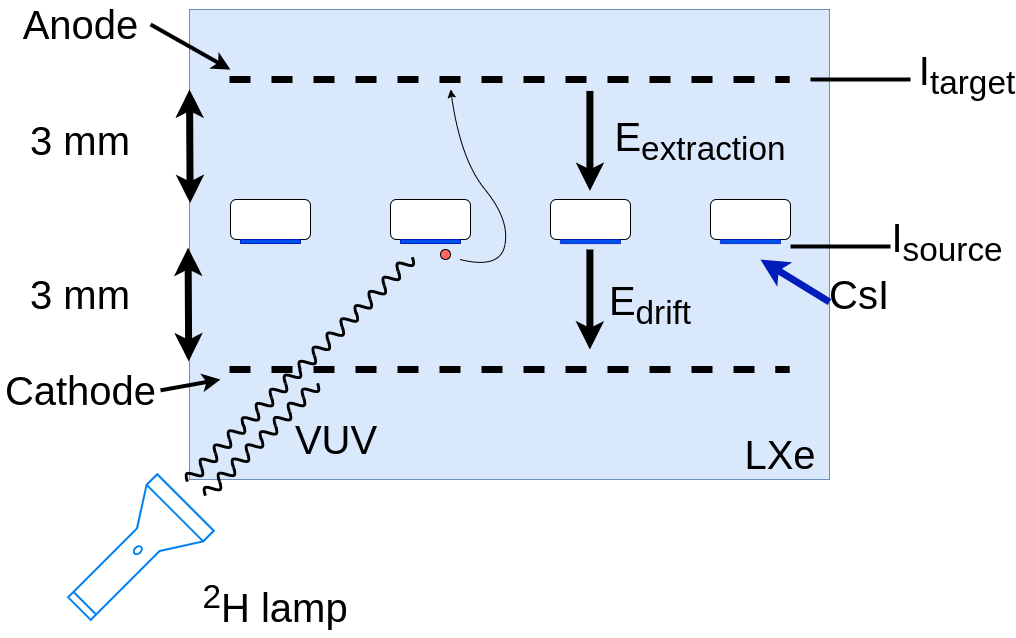}
    \caption{Schematic diagram of the setup used for the \ETEpe measurement.}
    \label{fig:s1_setup}
\end{figure}

Figure \ref{fig:s1_setup} shows the setup used for the measurement of the transfer efficiency of \sone-induced photoelectrons produced on the THGEM photocathode. In this arrangement, a THGEM coated with a 450~nm-thick CsI layer on its bottom face was positioned 3~mm above a mesh that functioned as the cathode, with a second mesh placed 3~mm above the THGEM serving as the anode.

A deuterium VUV lamp (Hammamatsu L879-01) was used to illuminate the CsI-coated face of the THGEM through the side viewport.
The lamp was mounted in an enclosure flushed with \ntwo to avoid VUV absorption in air, providing a continuous spectrum above the cutoff wavelength of the window ($\sim170$~nm).
This resulted in the release of photoelectrons from the photocathode, with their subsequent focusing into the THGEM holes, under the voltage applied across the electrode.
To prevent the electrons from landing on the top face of the THGEM, a strong electric field (\eextraction) was applied above the electrode, forcing the electrons upwards towards the mesh, where they were collected.
A moderate electric field below the THGEM (\edrift) replicated typical conditions of a realistic TPC.

\changes{
The stability of the lamp was measured both independently and \textit{in situ} and no significant variations were observed.
In order to minimize the effect of the charging-up of the THGEM, the measurements were separated by 10-minute intervals in which the light source was blocked.
}

In this experiment, we investigated the photoelectron transfer efficiency under voltages across the THGEM (\dvthgem) in the range $1.5-2.25$~kV. For lower voltages, the resulting effective QE of the photocathode was not high enough, and for higher values the required voltages became impractical. We explored drift field values up to 1.76~kV/cm and extraction fields up to 5.67~kV/cm.

\begin{figure}
    \centering
    \includegraphics[width=0.49\columnwidth]{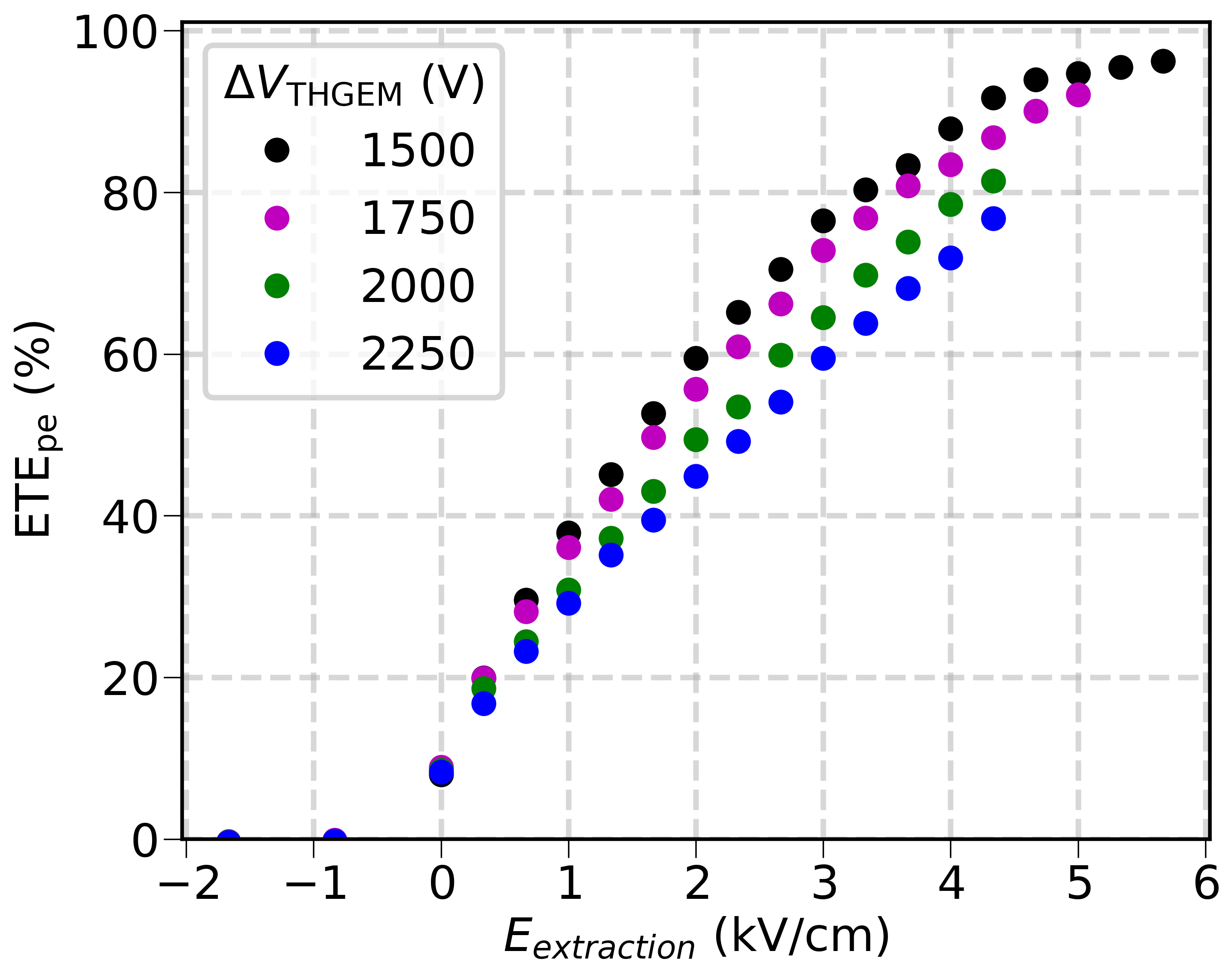}
    \includegraphics[width=0.49\columnwidth]{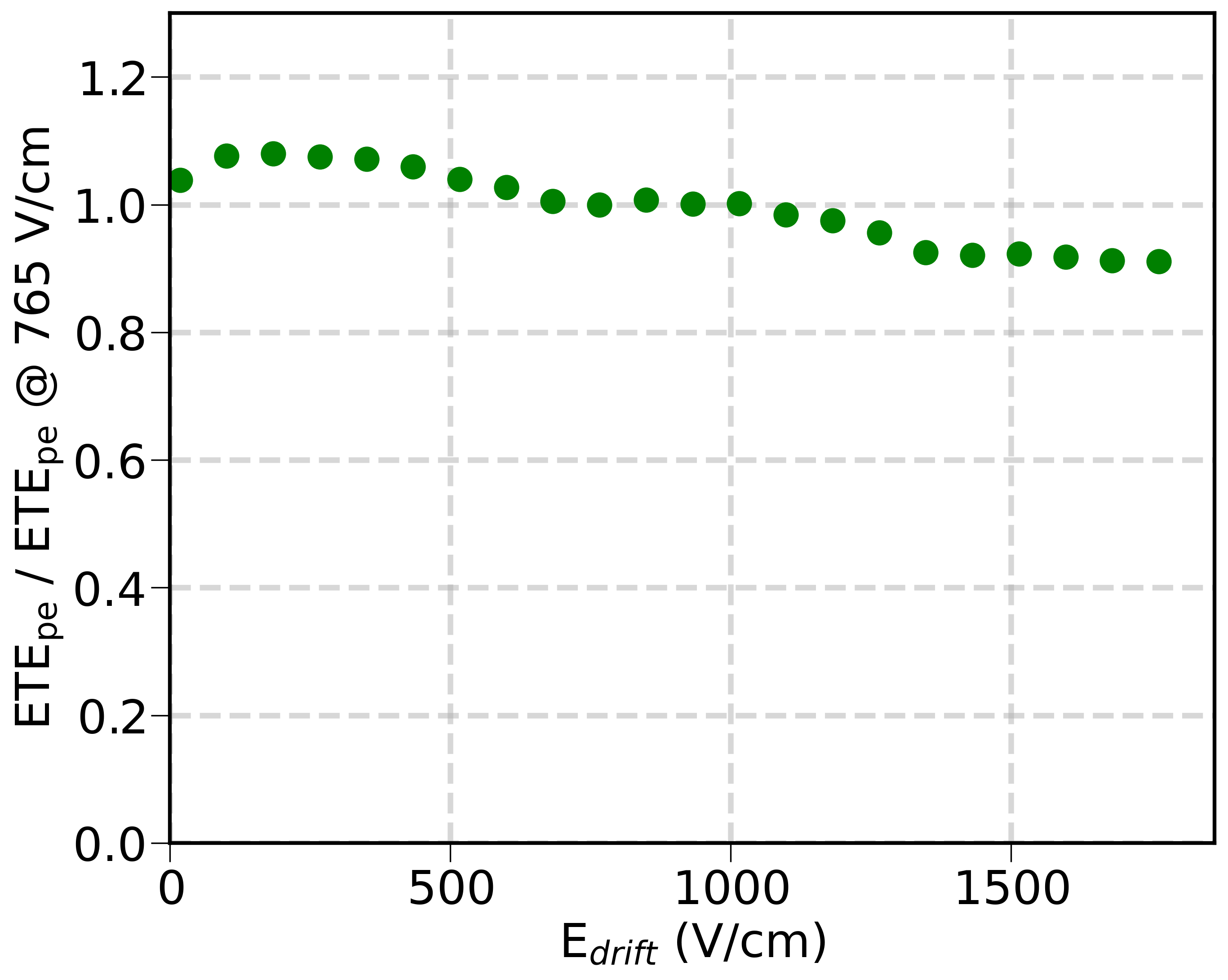}
    \caption{Left: Measured ETE for photoelectrons as a function of \eextraction for different values of \dvthgem and with \edrift $\approx760$~V/cm. Right: Measured ETE for photoelectrons as a function of the drift field for $\dvthgem=2.0$~kV and $\eextraction=4.0$~kV/cm normalized to the value obtained at $\edrift = 765$~V/cm.}
    \label{fig:s1_ete}
\end{figure}

\begin{figure}
    \centering
    \includegraphics[width=0.7\columnwidth]{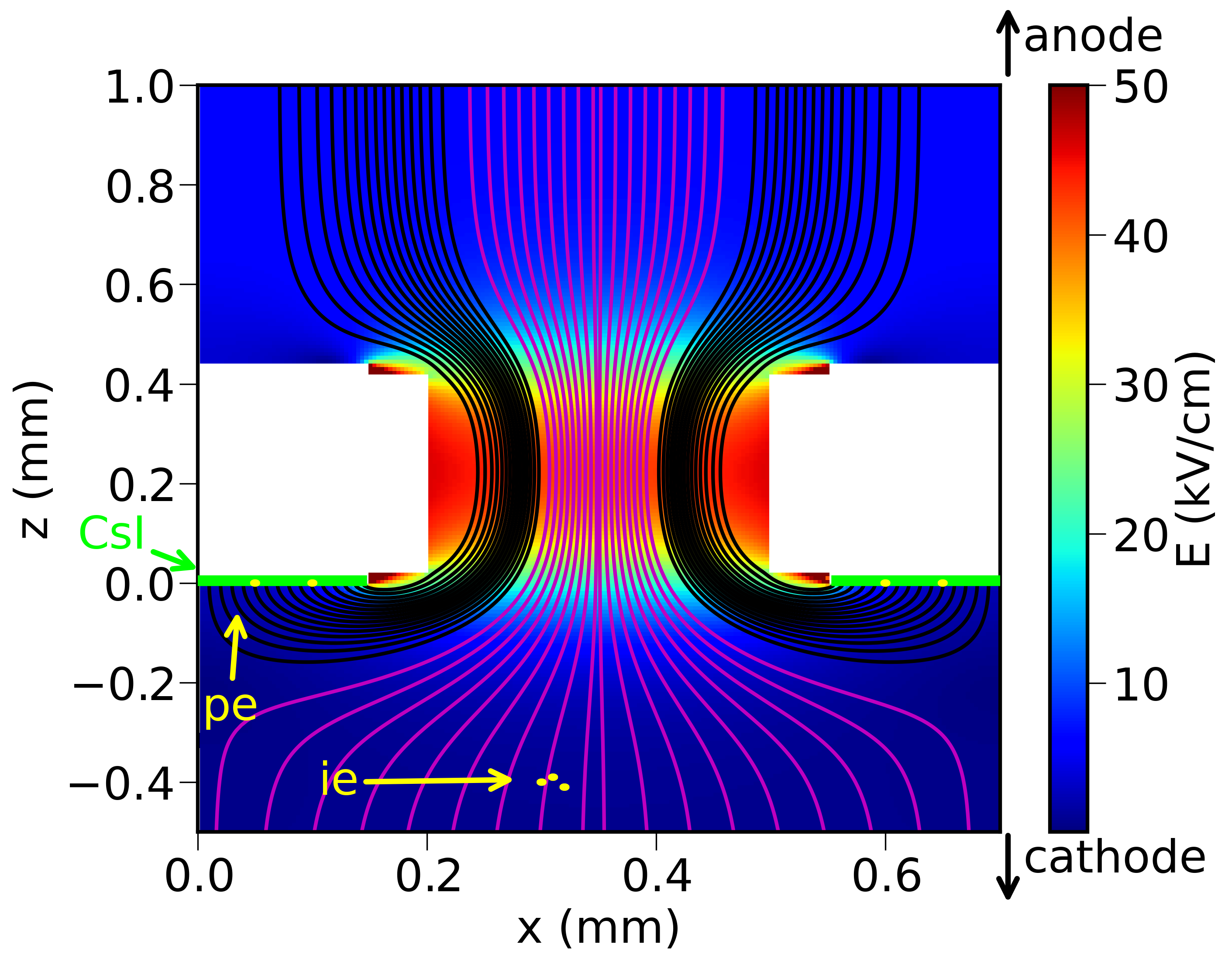}
    \caption{Field lines for \sone-induced photoelectrons (black) and ionization electrons (magenta) simulated in COMSOL Mutiphysics\registered. The THGEM surface is shown in white and the CsI photocathode in green. The background color denotes the intensity of the electric field. Field lines start at the photocathode and end at the anode (not shown). The voltages in this simulation were set such that $\edrift \approx 600$~V/cm, $\dvthgem=2.0$~kV, and $\eextraction=6.0$ kV/cm. The high curvature and the proximity to the insulator surface of the photoelectrons' field lines (a proxy for their trajectory) have an impact on the transfer efficiency.}
    \label{fig:s1p_field_lines}
\end{figure}

Figure \ref{fig:s1_ete} shows the measured ETE of \sone-induced photoelectrons (\ETEpe). The left panel displays the measured \ETEpe for different values of \dvthgem as a function of the extraction field for drift fields in the range $740-780$~V/cm (depending on \dvthgem). We measured an \ETEpe above 90\% for $\dvthgem = 1.5$~kV and $\dvthgem = 1.75$~kV. However, for $\dvthgem = 2.0-2.25$~kV, voltage limitations (discharges, probably from connections not fully immersed in LXe) impeded complete measurements. In these cases, we report an \ETEpe close to or above 80\% \original{but a sensible extrapolation of the data indicates that an \ETEpe above 90\% can also be achieved for higher extraction fields}. The reduction in \ETEpe at higher \dvthgem values can be attributed to the increased curvature of the field lines near the holes.
This increased curvature results in poorer electron focusing into the holes and less efficient extraction of electrons towards the anode. Simulations performed in COMSOL Multiphysics\registered, support this hypothesis.
Figure \ref{fig:s1p_field_lines} displays an example of the field lines (a proxy for the trajectory) for photoelectrons (black) and ionization electrons (magenta) for a specific set of voltages. While ionization electrons traverse the structure close to the center of the hole, the photoelectrons' field lines approach the surface of the insulator, which can lead to photoelectron losses.

The right panel of Figure \ref{fig:s1_ete} shows the \ETEpe as a function of the drift field for $\dvthgem=2.0$~kV and $\eextraction=4.0$~kV/cm normalized to the \ETEpe measured at $\edrift = 765$~V/cm. We observe that the \ETEpe deteriorates as the drift field increases. Our simulations indicate that this effect can be attributed to a reduced focusing efficiency of photoelectrons into the THGEM holes at higher fields.

\changes{
The slight decrease in \ETEpe with \dvthgem for a fixed extraction field is attributed to electron loss to the top face of the THGEM.
For higher values of \dvthgem, a larger fraction of field lines are attracted to the top face instead of the mesh.
This hypothesis is supported by COMSOL simulations.
}

\subsection{Ionization-electron transfer efficiency}
\label{sec:ete_s2}

\begin{figure}
    \centering
    \includegraphics[width=0.7\columnwidth]{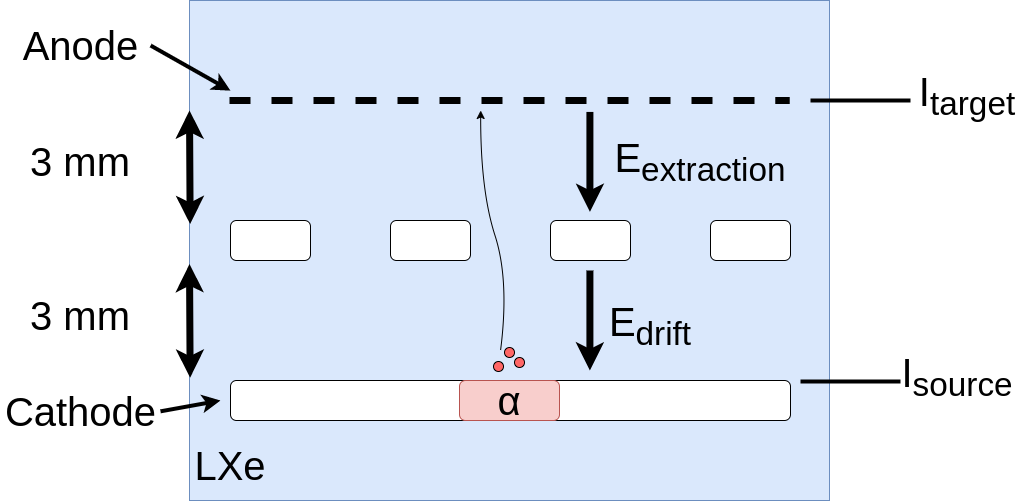}
    \caption{Schematic diagram of the setup used for the ionization-electron ETE measurement.}
    \label{fig:s2_setup}
\end{figure}

The setup used for the radiation-induced ionization-electron measurement is depicted in Figure~\ref{fig:s2_setup}.
A 1/2" diameter \am source with an activity of 10 $\mu$Ci was used to produce the ionization electrons. The source was mounted on a flat stainless steel frame that acted as a cathode. The surface of the source was leveled with the frame to ensure the uniformity of the electric field. The frame was mounted 3~mm below an uncoated THGEM. A mesh acting as an anode was mounted 3~mm above the THGEM.

In this case, the electrons induced by $\alpha$-particle interactions in LXe drifted towards the THGEM and were focused into its holes by applying a voltage across the electrode. They were then pulled to the top mesh by an extraction field.

\changes{
The contribution of photoelectron emission from the cathode and bottom-THGEM surfaces is estimated as negligible.
Assuming $\Wsc = 17.9$~eV in LXe \cite{Doke_2002} and a source $\alpha$-emission rate of $1.9\cdot10^5$ Hz, we estimate a rate of $5.8 \cdot 10^{10}$ photons/s.
Of those, roughly half go towards each electrode.
The QE of stainless steel and gold is $\lesssim 10^{-3}$ \cite{10.1063/5.0059497} and at the relevant drift fields a large fraction of the emitted photoelectrons backscatter to the cathode \cite{APRILE1994328}, so the expected contribution to the current from photoemission is $\lesssim 1$~pA.
This is roughly two orders of magnitudes smaller than the currents measured in our experiments.
}
\changes{
Furthermore, in order to determine if the charging up of the THGEM had an impact on the results we repeated the measurements in a different sequence and we scanned the voltages applied both in ascending and descending order.
All datasets were in reasonable agreement.
}

Similar to the \ETEpe measurement, \dvthgem ranged from 1.5~kV to 2.25~kV, \edrift up to 1.67~kV/cm, and \eextraction up to 6~kV/cm.

\begin{figure}
    \centering
    \includegraphics[width=0.49\textwidth]{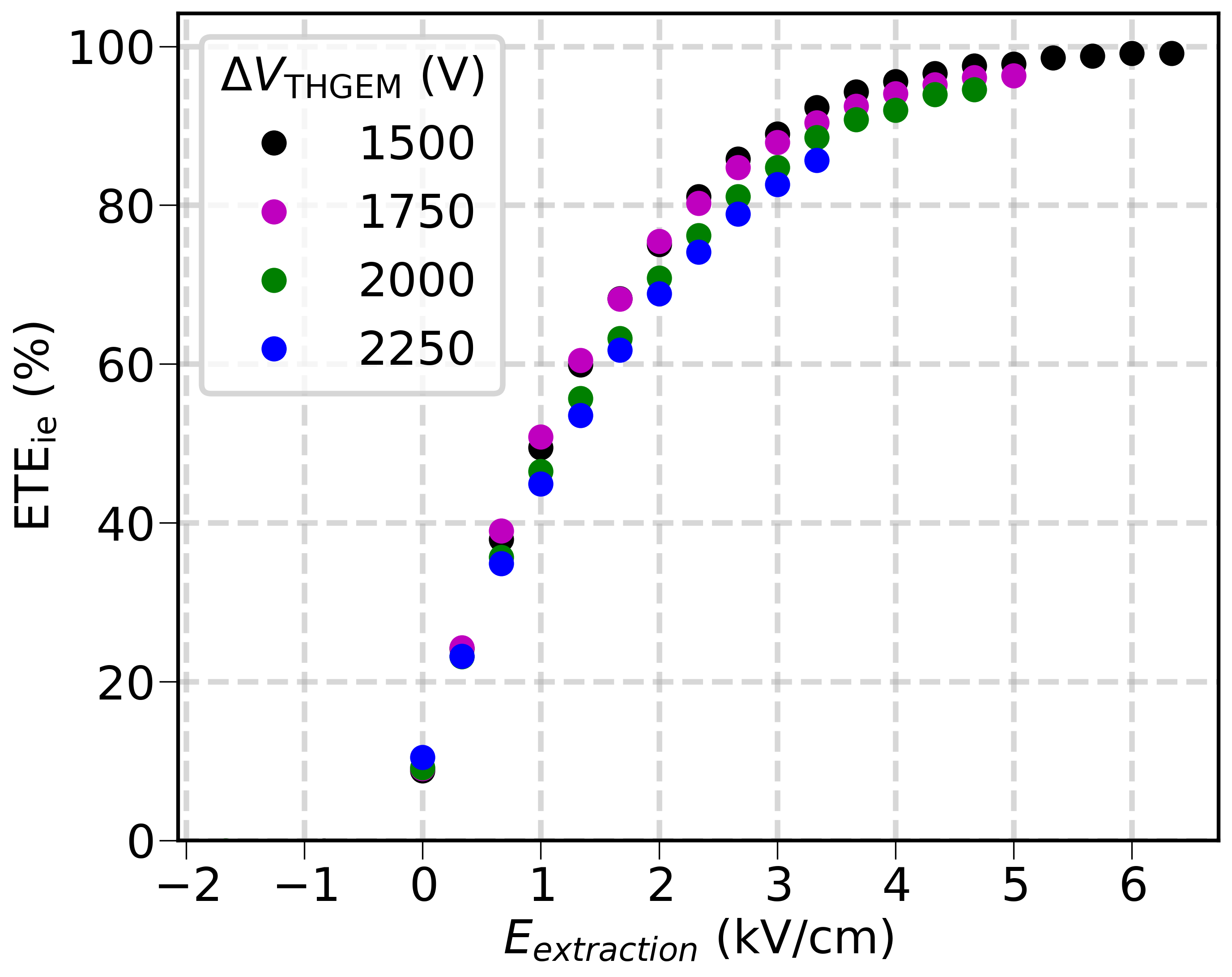}
    \includegraphics[width=0.49\textwidth]{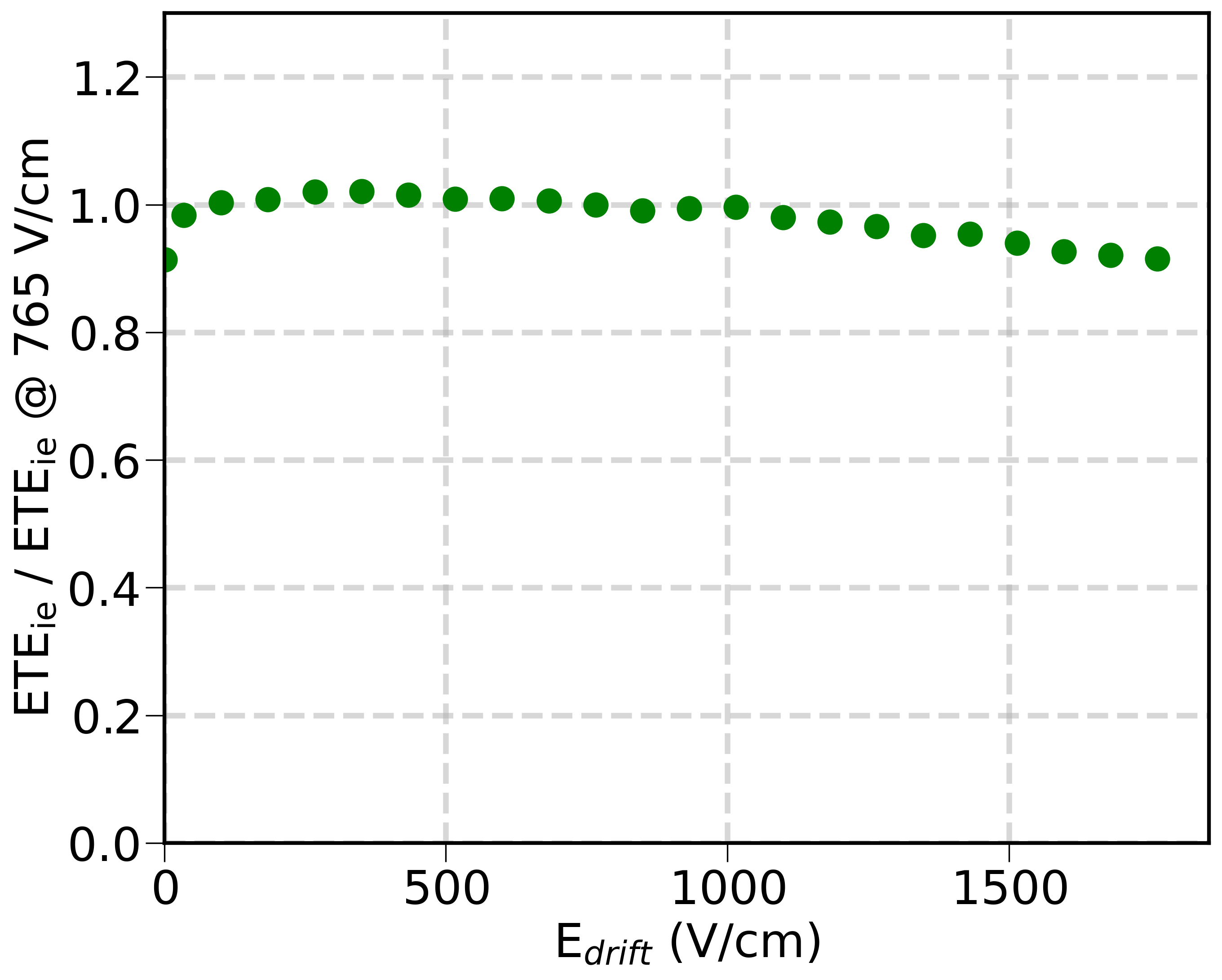}
    \caption{Left: Measured ETE for ionization electrons as a function of \eextraction for different values of \dvthgem and \edrift $\approx$ 760 V/cm.
    Right: Measured ETE for ionization electrons as a function of the drift field for $\dvthgem = 2.0$ kV and $\eextraction = 4.0$ kV/cm normalized to the value obtained at $\edrift = 765$~V/cm.}
    \label{fig:s2_ete}
\end{figure}

The measured ETE for ionization electrons (\ETEie) is displayed in Figure \ref{fig:s2_ete}. The left panel shows the \ETEie as a function of \eextraction for different values of \dvthgem. We measure an \ETEie above 90\% for $\dvthgem\leq2.0$~kV. For $\dvthgem = 2.25$~kV, voltage limitations impeded a complete measurement\original{, but a sensible extrapolation of the data indicates that a similar result can be achieved}. In the right panel, we present the \ETEie as a function of \edrift for $\dvthgem = 2$~kV and $\eextraction = 4.0$~kV/cm normalized to the \ETEpe measured at \edrift = 765 V/cm. We do not observe a significant deterioration of the \ETEie for higher fields. The gradual decline can be attributed to the decrease in the focusing efficiency of ionization electrons into the THGEM holes, which was confirmed by COMSOL simulations.

\changes{
As for the pe measurement, the slight decrease in \ETEie with \dvthgem for a fixed extraction field is attributed to electron loss to the top face of the THGEM following the same reasoning given in Section \ref{sec:ete_s1prime}.
} 

\section{Simulation of electron transport}
\label{sec:simulations}

In order to validate the results described in Section \ref{sec:ete}, we performed a number of detailed electron transport simulations. First, and as a reference, we considered the simplest electron transport case: electrons following the field lines without diffusion. A large number of field lines were calculated in COMSOL, starting from either the bottom surface of the THGEM (for photoelectrons) or from the cathode (for ionization electrons). Counting the number of field lines ending on the mesh electrode (anode) provided a rough estimate of the ETE in each configuration.

\changes{
However, diffusion is expected to play a significant role in the propagation of electrons through the THGEM.
For a transverse diffusion coefficient in the range $50-90$~cm$^2$/s, we expect a standard deviation of the electron trajectories in the range $45-60$~$\mu$m, which is not negligible compared to the radius of the THGEM holes (150~$\mu$m).
}
\original{In a second simulation performed}
\changes{Therefore, a second simulation was performed}
in Garfield++ \cite{garfield}, in which electrons were transported microscopically in discrete steps in a geometry equivalent to that of each measurement. The propagation of electrons was performed according to the electric field configuration obtained from COMSOL simulations of the given geometry and the atomic collision probabilities provided by the NEST package \cite{Szydagis_2011,szydagis_m_2023_7577399}. 10$^4$ electrons were generated uniformly on the bottom surface of the THGEM (for photoelectrons) or on the cathode (for ionization electrons). Each electron was tracked until it left the drift medium (i.e. when it reached a conducting plane or the FR4 insulator). Counting the number of electrons reaching the mesh electrode yielded an estimate of the ETE, mimicking the experiment. The maximum step size was set to 10~$\mu$m, although we did not observe any significant difference in our results with slightly larger or smaller steps.

\begin{figure}
    \centering
    \includegraphics[width=0.7\columnwidth]{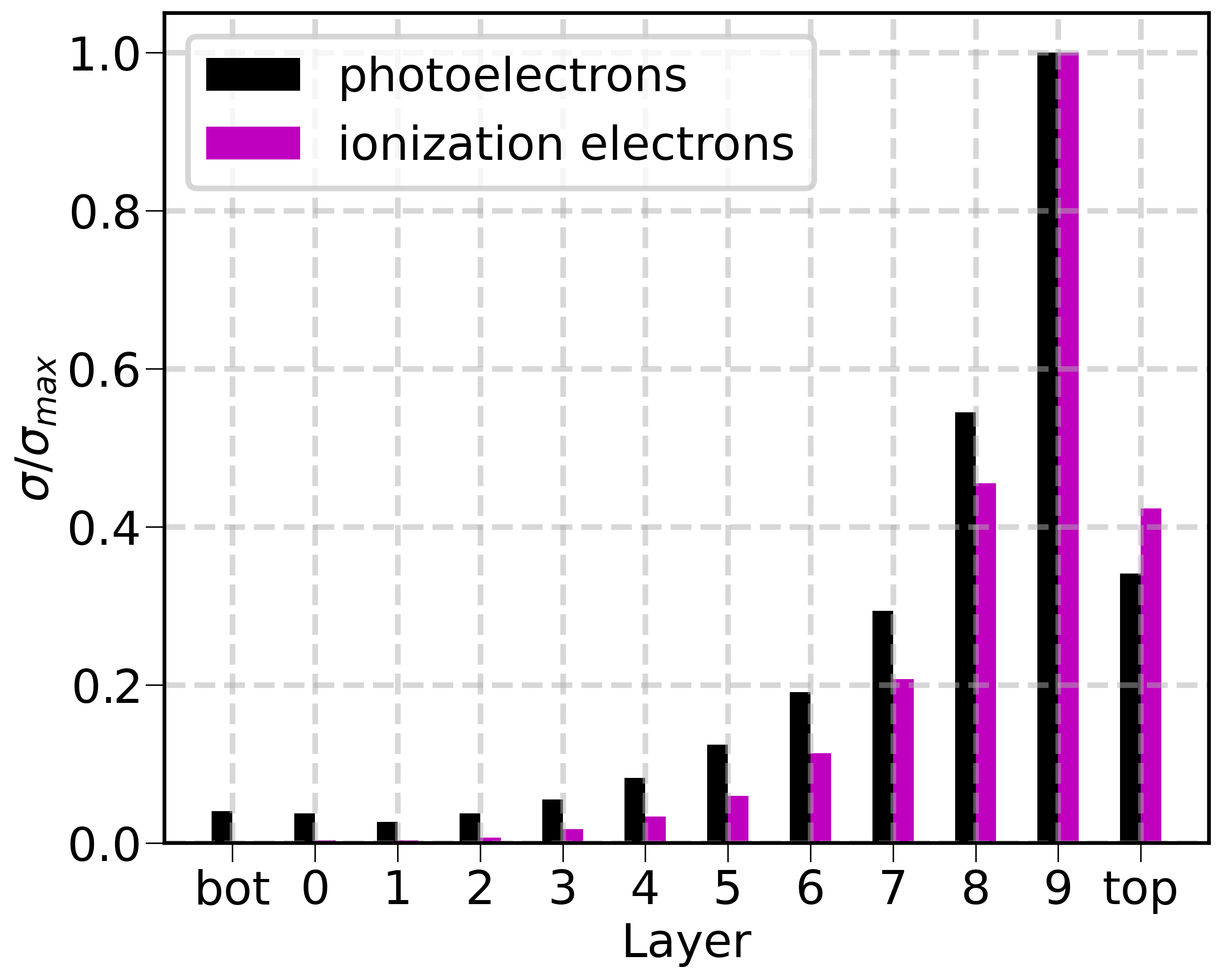}
    \caption{Normalized distribution of the surface charge density obtained for photoelectrons and ionization electrons under $\dvthgem=2$~kV and $\edrift=760$~V/cm. The inner wall of the THGEM holes was divided into 10 layers of equal thickness and numbered from bottom to top. "bot" and "top" indicate the surface of the hole rim. The resulting charge density is the average of the values obtained for all extraction fields.}
    \label{fig:surface_charge_densities}
\end{figure}

Due to the high currents needed to perform the measurements ($\gtrsim100$~pA), loss of electrons on the walls of the insulator can lead to significant surface charge densities that modify the local field distribution inside the hole and can affect the transfer efficiency of electrons. As a rough approximation, the simulations were performed in two stages. First, we obtained a first field map assuming a null surface charge density on the exposed surfaces of the insulator. With this map, we ran the electron simulation and estimated the distribution of the surface charge density ($\sigma$) by counting the number of electrons lost in different $z$-slices of the insulator. The value of $\sigma$ was obtained for 12 different surfaces: the top and bottom rims and ten slices of the inner wall of the hole in the insulator. Figure \ref{fig:surface_charge_densities} shows the relative surface charge density ($\sigma/\sigma_{max}$) in each surface for photoelectrons and ionization electrons. As a second stage, we implemented the surface charge density distribution calculated from the electron ``landing sites'' in COMSOL to retrieve an updated field map. In this calculation, the absolute value of $\sigma_{max}$ was a free parameter and we therefore scanned a range of values to obtain the best match to the data. Finally, the updated field map was used to run a second electron transport simulation giving the final result.

\begin{figure}
    \centering
    \includegraphics[width=0.49\textwidth]{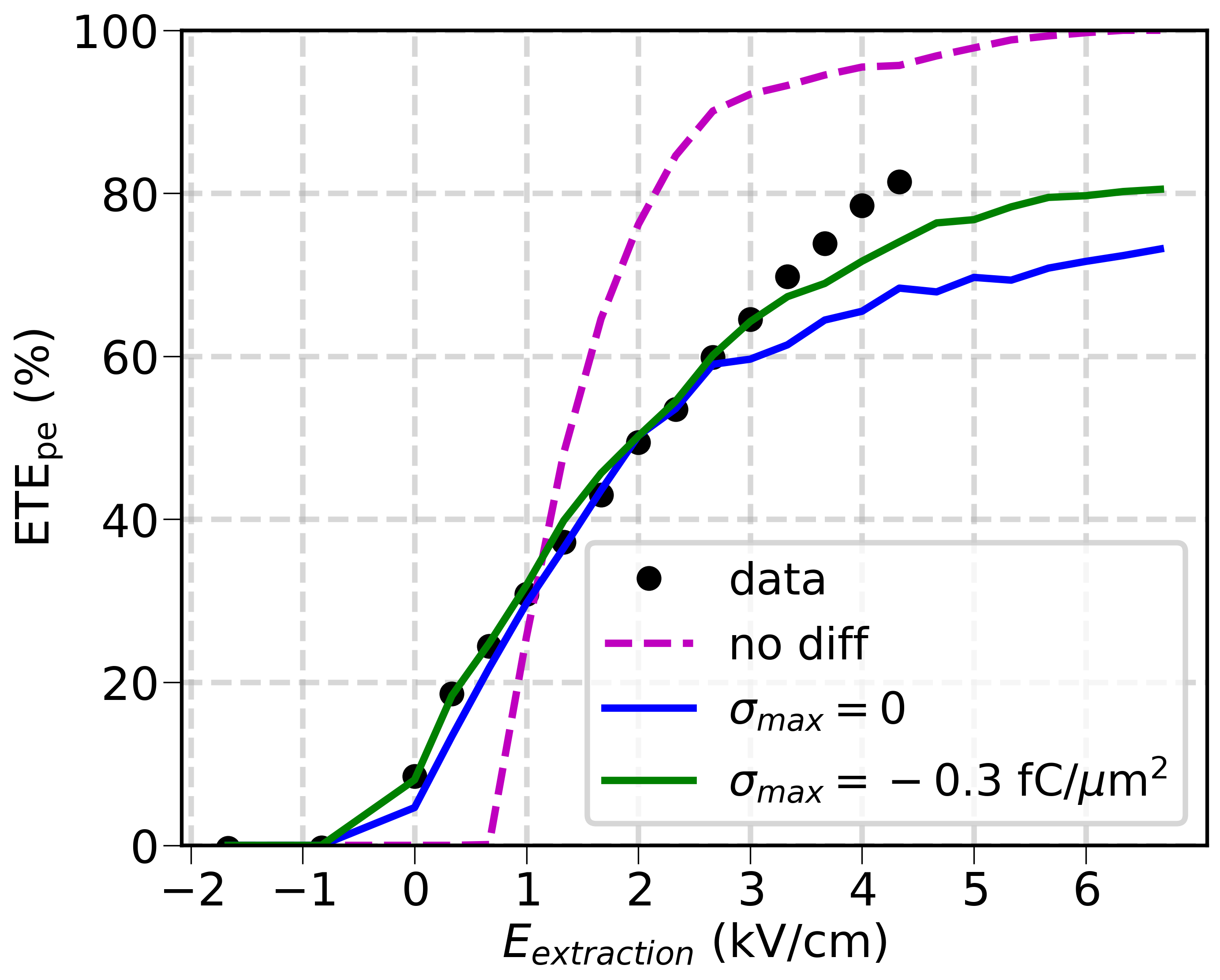}
    \includegraphics[width=0.49\textwidth]{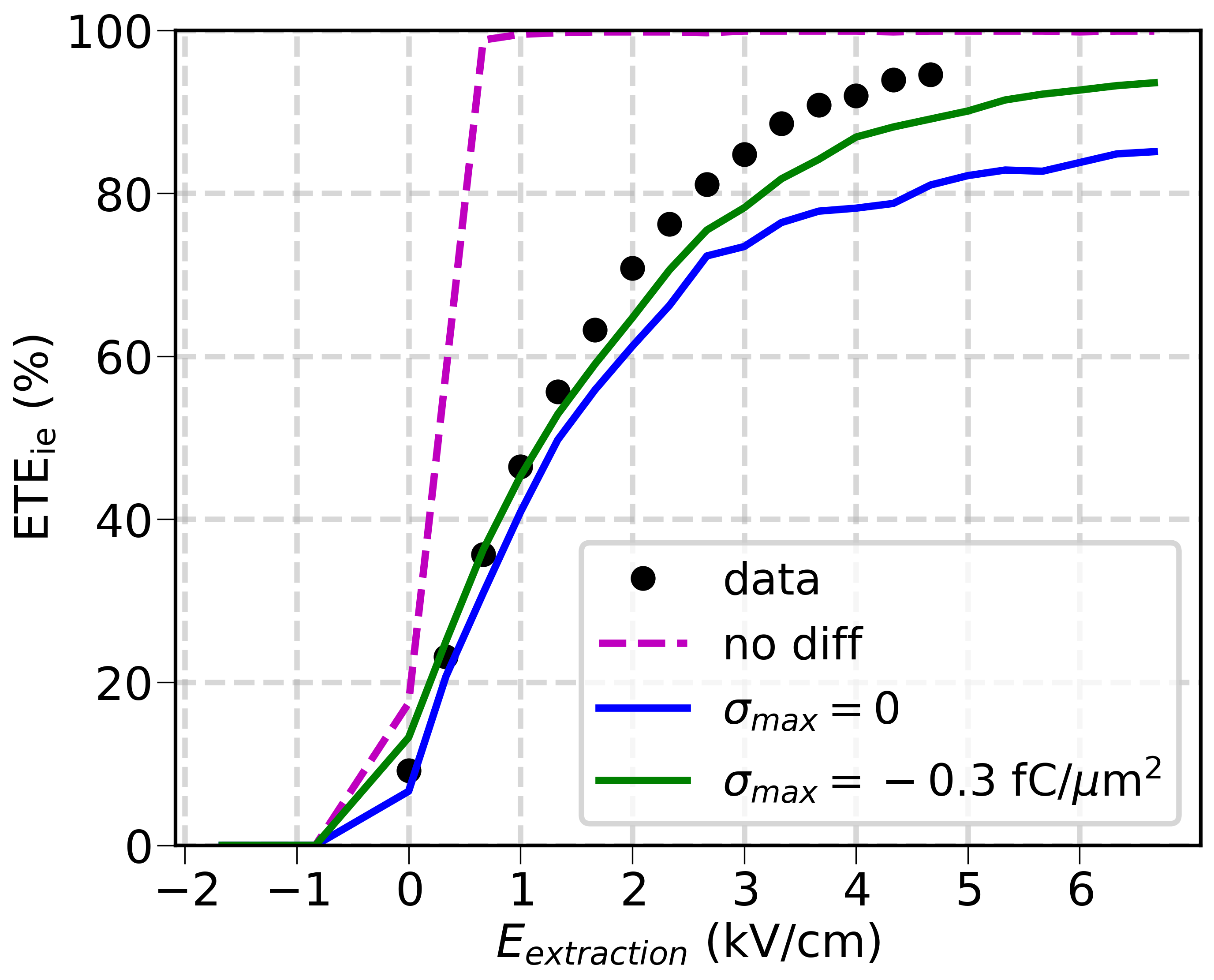}
    \caption{Comparison between data (black circles) and simulation for the ETE of photoelectrons (left panel) and ionization electrons (right panel). The dashed magenta line indicates the ETE predicted by a simulation with no diffusion, while the solid lines represent the full microscopic electron transport with different surface charge densities. The latter matches the data much better, assuming the surface charge density distribution of Figure \ref{fig:surface_charge_densities} and the values of $\sigma_{max}$ displayed in the legend.}
    \label{fig:data_sim_comparison}
\end{figure}

The simulations were run with $\dvthgem=2$~kV, $\edrift=760$~V/cm, and the same extraction fields as for the experimental data.
Figure \ref{fig:data_sim_comparison} shows the comparison between data (black circles) and simulation (solid lines), accounting for different values of the surface charge density for photoelectrons (left panel) and ionization electrons (right panel). The no-diffusion simulation is displayed as a dashed magenta line for reference.

We observe that the no-diffusion case does not reproduce data sufficiently well. A much better agreement was found after implementing electron diffusion in the simulation even without applying a surface charge density to the insulator. The simulations with $\sigma_{max} = -0.3$~fC/$\mu$m$^2$ further improve the agreement between data and simulations, although it is still not perfect. Possible reasons for the disparity between the results include the oversimplification in estimating and implementing the surface charge density distribution, a time-dependent charging-up process of the THGEM \cite{Pitt_2018} that is not reflected in the simulation, and a voltage dependence on the spatial distribution of $\sigma$. Moreover, in our approach, we abruptly transitioned from a zero charge density to a significantly high value in a single step, which may have contributed to the discrepancy. Employing an iterative process involving smaller incremental steps, where we gradually introduce lower surface charge densities to the field simulation, recompute the electron transport, and update the field simulation, is likely to yield a more precise outcome. By using sufficiently small incremental changes in the surface charge density, a more realistic estimation of the charging up of the THGEM might be achieved.
\changes{
However, these simulations are extremely costly in terms of computation time and a detailed description of the charging-up of the THGEM is beyond the scope of this article.
}

\changes{
Note that the no-diffusion case underestimates the \ETEpe for extraction fields below 1~kV/cm.
By analyzing the field line distribution, we observe that there are some drift regions where electrons are absorbed (opaque) and other drift regions where electrons are transmitted (transparent).
We attribute this effect to electrons diffusing from opaque to transparent drift regions.
}
\section{Discussion and conclusions}
\label{sec:discussion}

\begin{figure}
    \centering
    \includegraphics[width=0.7\textwidth]{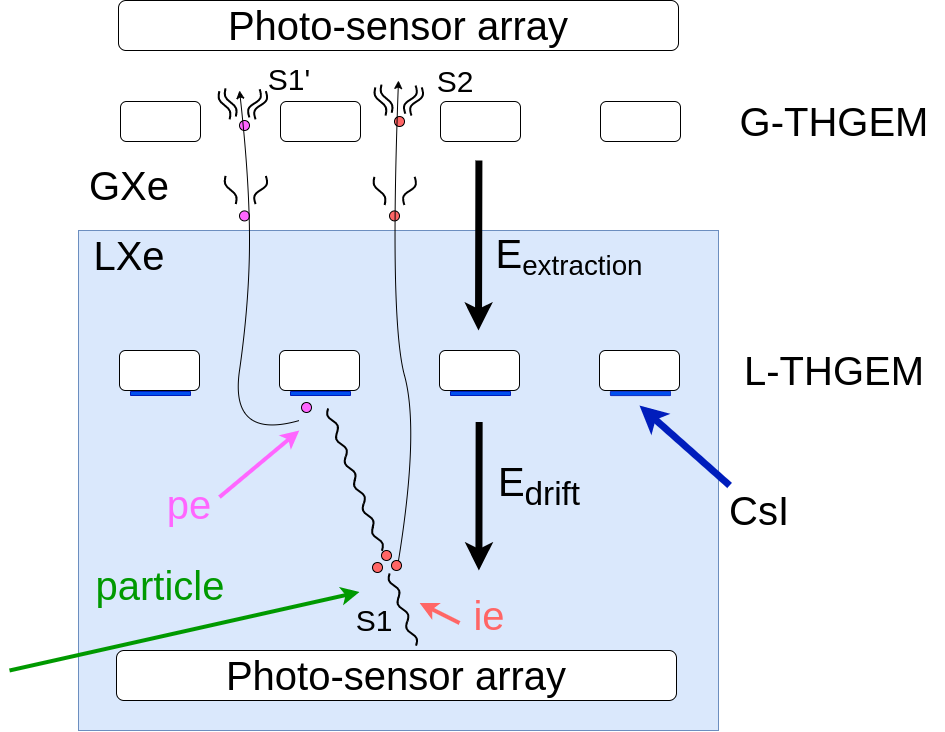}
    \caption{Schematic design for the ``cascaded-LHM''. A CsI-coated THGEM is immersed in LXe (L-THGEM) and an uncoated THGEM is placed in the gas phase (G-THGEM). Photosensor arrays detect the primary scintillation and the secondary light pulses produced in the G-THGEM holes (see text for details).}
    \label{fig:dual_phase_concept}
\end{figure}

The measurements presented here indicate that a high transfer efficiency, close to unity, across a THGEM immersed in LXe can be achieved for both photoelectrons and ionization electrons under the same electric field configurations with field strengths adequate for the operation of single- or dual-phase xenon detectors.
This encourages the development of detectors based on an immersed electrode coated with CsI for improved \sone detection. Moreover, this idea can be applied to a number of detector concepts and is suitable for both single-phase and dual-phase detectors.

As proposed in \cite{Breskin_2022}, one possibility is to couple a micro-strip plate to the THGEM and produce a small charge avalanche and EL directly in the liquid. Another option is to implement the micro-strip pattern directly on the back (here, top face) of the THGEM. This would be a simpler scheme that would not need an intense extraction field to avoid electron loss, as the electrons are collected on the back side directly.

For dual-phase detectors, the interface can lie above the THGEM with a mesh or another THGEM above. The top face of the THGEM immersed in liquid would act as a gate electrode like in traditional dual-phase detectors.
Figure \ref{fig:dual_phase_concept} shows a conceptual design for the ``cascaded-LHM'': a dual-phase LXe TPC with an immersed CsI-coated THGEM (L-THGEM) and an uncoated THGEM in the gas phase (G-THGEM). In this concept, primary scintillation photons can be detected directly by a photosensor array at the TPC bottom (\sone), and they also extract photoelectrons from the photocathode. These photoelectrons are focused into the L-THGEM holes and extracted to the gas phase by an intense field. The extracted photoelectrons are subsequently focused into the G-THGEM holes where they produce a large EL pulse labeled \soneprime. Ionization electrons produced in the interaction are also transferred across the L-THGEM, extracted to the gas, and focused into the G-THGEM holes to produce a second light pulse (\stwo).

The large \soneprime pulses allow for the unambiguous detection of single-photon \sone signals, as they cannot be mimicked by the dark counts of the photosensors. \changes{A photoelectron extracted from CsI that is successfully transferred through the THGEM will produce an unambiguous signal with a high signal-to-noise ratio. In the single-phase case, a light yield of $\sim$300 photons/electron using 10~$\mu$m wires at 6~kV was reported in \cite{Aprile_2014}. For a typical light collection efficiency of $\sim$40\% and a photosensor QE of $\sim$35\%, the S2 gain would be $\sim$40~pe/electron. For dual-phase detectors, the XENON1T experiment reports $\sim$12~pe/electron \cite{XENON:2017lvq}.
Thus, regardless of whether the amplification occurs in liquid or in gas, the signal-to-noise ratio is large enough for dark counts not to be a concern, even for photosensors with high dark count rates such as SiPMs, or even CMOS-SPADs.
}

Our study focused on a specific perforated electrode, namely a THGEM with 0.3~mm diameter holes and 0.7~mm pitch.
Further studies need to be carried out with other electrodes, which might be better suited for the detector concepts discussed here. Based on previous works \cite{Erdal:2017yiu}, a standard Gas Electron Multiplier (GEM) or a single-conical (single-mask) GEM can outperform THGEMs. These thinner structures can reach more intense electric fields on the surface, which has a direct impact on the QE of CsI, and operate at lower voltages. On the other hand, various GEM transparencies and hole sizes need to be investigated to extract the full potential of this technique.

\appendix

\acknowledgments
The authors would like to thank Dr Sergei Shchemelinin from the Unit of Nuclear Engineering of the Ben-Gurion University of the Negev for his assistance with the CsI evaporation. The authors would also like to thank Dr. Eran Erdal and Dr. David Vartsky for their helpful discussions related to the experiment and their useful comments in preparing this manuscript.

\bibliographystyle{JHEP}
\bibliography{src/bibliography}

\providecommand{\href}[2]{#2}\begingroup\raggedright\begin{thebibliography}{10}

\bibitem{PhysRevD.98.102006}
{\scshape DarkSide Collaboration} collaboration, \emph{Darkside-50 532-day dark
  matter search with low-radioactivity argon},
  \href{https://doi.org/10.1103/PhysRevD.98.102006}{\emph{Phys. Rev. D}
  {\bfseries 98} (2018) 102006}.

\bibitem{PhysRevD.96.022008}
{\scshape XENON Collaboration} collaboration, \emph{Search for wimp inelastic
  scattering off xenon nuclei with xenon100},
  \href{https://doi.org/10.1103/PhysRevD.96.022008}{\emph{Phys. Rev. D}
  {\bfseries 96} (2017) 022008}.

\bibitem{PhysRevLett.123.251801}
{\scshape XENON Collaboration} collaboration, \emph{Light dark matter search
  with ionization signals in xenon1t},
  \href{https://doi.org/10.1103/PhysRevLett.123.251801}{\emph{Phys. Rev. Lett.}
  {\bfseries 123} (2019) 251801}.

\bibitem{PhysRevD.102.072004}
{\scshape XENON Collaboration} collaboration, \emph{Excess electronic recoil
  events in xenon1t},
  \href{https://doi.org/10.1103/PhysRevD.102.072004}{\emph{Phys. Rev. D}
  {\bfseries 102} (2020) 072004}.

\bibitem{PhysRevD.98.062005}
{\scshape LUX Collaboration} collaboration, \emph{Search for annual and diurnal
  rate modulations in the lux experiment},
  \href{https://doi.org/10.1103/PhysRevD.98.062005}{\emph{Phys. Rev. D}
  {\bfseries 98} (2018) 062005}.

\bibitem{PhysRevLett.122.131301}
{\scshape LUX Collaboration} collaboration, \emph{Results of a search for
  sub-gev dark matter using 2013 lux data},
  \href{https://doi.org/10.1103/PhysRevLett.122.131301}{\emph{Phys. Rev. Lett.}
  {\bfseries 122} (2019) 131301}.

\bibitem{YUAN2022137254}
Y.~Yuan, A.~Abdukerim, Z.~Bo, W.~Chen, X.~Chen, Y.~Chen et~al., \emph{A search
  for two-component majorana dark matter in a simplified model using the full
  exposure data of pandax-ii experiment},
  \href{https://doi.org/https://doi.org/10.1016/j.physletb.2022.137254}{\emph{Physics
  Letters B} {\bfseries 832} (2022) 137254}.

\bibitem{PhysRevLett.129.161805}
{\scshape XENON Collaboration} collaboration, \emph{Search for new physics in
  electronic recoil data from xenonnt},
  \href{https://doi.org/10.1103/PhysRevLett.129.161805}{\emph{Phys. Rev. Lett.}
  {\bfseries 129} (2022) 161805}.

\bibitem{2022arXiv220703764A}
J.~{Aalbers}, D.S.~{Akerib}, C.W.~{Akerlof}, A.K.~{Al Musalhi}, F.~{Alder},
  A.~{Alqahtani} et~al., \emph{{First Dark Matter Search Results from the
  LUX-ZEPLIN (LZ) Experiment}}, {\emph{arXiv e-prints} (2022) arXiv:2207.03764}
  [\href{https://arxiv.org/abs/2207.03764}{{\ttfamily 2207.03764}}].

\bibitem{PhysRevLett.129.161803}
{\scshape PandaX Collaboration} collaboration, \emph{First search for the
  absorption of fermionic dark matter with the pandax-4t experiment},
  \href{https://doi.org/10.1103/PhysRevLett.129.161803}{\emph{Phys. Rev. Lett.}
  {\bfseries 129} (2022) 161803}.

\bibitem{PhysRevLett.129.161804}
{\scshape PandaX Collaboration} collaboration, \emph{Search for light fermionic
  dark matter absorption on electrons in pandax-4t},
  \href{https://doi.org/10.1103/PhysRevLett.129.161804}{\emph{Phys. Rev. Lett.}
  {\bfseries 129} (2022) 161804}.

\bibitem{Aalbers_2016}
J.~Aalbers, F.~Agostini, M.~Alfonsi, F.~Amaro, C.~Amsler, E.~Aprile et~al.,
  \emph{Darwin: towards the ultimate dark matter detector},
  \href{https://doi.org/10.1088/1475-7516/2016/11/017}{\emph{Journal of
  Cosmology and Astroparticle Physics} {\bfseries 2016} (2016) 017}.

\bibitem{argo}
``Future dark matter searches with low-radioactivity argon.''
  \url{https://indico.cern.ch/event/765096/contributions/3295671/attachments/1785196/2906164/DarkSide-Argo\_ESPP\_Dec\_17\_2017.pdf}.

\bibitem{Aalseth_2018}
C.E.~Aalseth, F.~Acerbi, P.~Agnes, I.F.M.~Albuquerque, T.~Alexander, A.~Alici
  et~al., \emph{{DarkSide}-20k: A 20 tonne two-phase {LAr} {TPC} for direct
  dark matter detection at {LNGS}},
  \href{https://doi.org/10.1140/epjp/i2018-11973-4}{\emph{The European Physical
  Journal Plus} {\bfseries 133} (2018) }.

\bibitem{ANFIMOV2021165162}
N.~Anfimov, D.~Fedoseev, A.~Rybnikov, A.~Selyunin, S.~Sokolov and A.~Sotnikov,
  \emph{Study of silicon photomultiplier performance at different
  temperatures},
  \href{https://doi.org/https://doi.org/10.1016/j.nima.2021.165162}{\emph{Nuclear
  Instruments and Methods in Physics Research Section A: Accelerators,
  Spectrometers, Detectors and Associated Equipment} {\bfseries 997} (2021)
  165162}.

\bibitem{LUX_single_e_emission2020}
D.S.~Akerib, S.~Alsum, H.M.~Ara\'ujo, X.~Bai, J.~Balajthy, A.~Baxter et~al.,
  \emph{Investigation of background electron emission in the lux detector},
  \href{https://doi.org/10.1103/PhysRevD.102.092004}{\emph{Phys. Rev. D}
  {\bfseries 102} (2020) 092004}.

\bibitem{Kopec_2021}
A.~Kopec, A.~Baxter, M.~Clark, R.~Lang, S.~Li, J.~Qin et~al., \emph{Correlated
  single- and few-electron backgrounds milliseconds after interactions in
  dual-phase liquid xenon time projection chambers},
  \href{https://doi.org/10.1088/1748-0221/16/07/P07014}{\emph{Journal of
  Instrumentation} {\bfseries 16} (2021) P07014}.

\bibitem{XENON1T_single_e_emission2022}
{\scshape XENON Collaboration} collaboration, \emph{Emission of single and few
  electrons in xenon1t and limits on light dark matter},
  \href{https://doi.org/10.1103/PhysRevD.106.022001}{\emph{Phys. Rev. D}
  {\bfseries 106} (2022) 022001}.

\bibitem{Aprile_2014}
E.~Aprile, H.~Contreras, L.W.~Goetzke, A.J.M.~Fernandez, M.~Messina,
  J.~Naganoma et~al., \emph{Measurements of proportional scintillation and
  electron multiplication in liquid xenon using thin wires},
  \href{https://doi.org/10.1088/1748-0221/9/11/P11012}{\emph{Journal of
  Instrumentation} {\bfseries 9} (2014) P11012}.

\bibitem{Giboni_2014}
K.L.~Giboni, X.~Ji, H.~Lin and T.~Ye, \emph{On dark matter detector concepts
  with large-area cryogenic gaseous photo multipliers},
  \href{https://doi.org/10.1088/1748-0221/9/02/C02021}{\emph{Journal of
  Instrumentation} {\bfseries 9} (2014) C02021}.

\bibitem{Lin_2021}
Q.~Lin, \emph{Proposal of a geiger-geometry single-phase liquid xenon time
  projection chamber as potential detector technique for dark matter direct
  search}, \href{https://doi.org/10.1088/1748-0221/16/08/P08011}{\emph{Journal
  of Instrumentation} {\bfseries 16} (2021) P08011}.

\bibitem{Juyal_2021_wires}
P.~Juyal, K.~Giboni, X.~Ji and J.~Liu, \emph{On the electrode configurations in
  a large single phase liquid xenon detector for dark matter searches},
  \href{https://doi.org/10.1088/1748-0221/16/08/P08028}{\emph{Journal of
  Instrumentation} {\bfseries 16} (2021) P08028}.

\bibitem{Kuger_2022}
F.~Kuger, J.~Dierle, H.~Fischer, M.~Schumann and F.~Toschi, \emph{Prospects of
  charge signal analyses in liquid xenon {TPCs} with proportional scintillation
  in the liquid phase},
  \href{https://doi.org/10.1088/1748-0221/17/03/p03027}{\emph{Journal of
  Instrumentation} {\bfseries 17} (2022) P03027}.

\bibitem{Wei_2022}
Y.~Wei, J.~Qi, E.~Shockley, H.~Xu and K.~Ni, \emph{Performance of a radial time
  projection chamber with electroluminescence in liquid xenon},
  \href{https://doi.org/10.1088/1748-0221/17/02/C02002}{\emph{Journal of
  Instrumentation} {\bfseries 17} (2022) C02002}.

\bibitem{Breskin_2013}
A.~Breskin, \emph{Liquid hole-multipliers: A potential concept for large
  single-phase noble-liquid tpcs of rare events},
  \href{https://doi.org/10.1088/1742-6596/460/1/012020}{\emph{Journal of
  Physics: Conference Series} {\bfseries 460} (2013) 012020}.

\bibitem{Arazi:2015uja}
L.~Arazi, E.~Erdal, A.E.C.~Coimbra, M.L.~Rappaport, D.~Vartsky, V.~Chepel
  et~al., \emph{{Liquid Hole Multipliers: bubble-assisted electroluminescence
  in liquid xenon}},
  \href{https://doi.org/10.1088/1748-0221/10/08/P08015}{\emph{JINST} {\bfseries
  10} (2015) P08015} [\href{https://arxiv.org/abs/1505.02316}{{\ttfamily
  1505.02316}}].

\bibitem{Arazi:2016nvi}
L.~Arazi, E.~Erdal, Y.~Korotinsky, M.~Rappaport, A.~Roy, S.~Shchemelinin
  et~al., \emph{{Progress with Xenon Liquid Hole Multipliers}},  in \emph{{2016
  IEEE Nuclear Science Symposium and Medical Imaging Conference}}, p.~8069864,
  2016, \href{https://doi.org/10.1109/NSSMIC.2016.8069864}{DOI}
  [\href{https://arxiv.org/abs/1708.00073}{{\ttfamily 1708.00073}}].

\bibitem{Erdal:2015kxa}
E.~Erdal, L.~Arazi, V.~Chepe, M.L.~Rappaport, D.~Vartsky and A.~Breskin,
  \emph{{Direct observation of bubble-assisted electroluminescence in liquid
  xenon}}, \href{https://doi.org/10.1088/1748-0221/10/11/P11002}{\emph{JINST}
  {\bfseries 10} (2015) P11002}
  [\href{https://arxiv.org/abs/1509.02354}{{\ttfamily 1509.02354}}].

\bibitem{Erdal:2016jhl}
E.~Erdal, L.~Arazi, M.L.~Rappaport, S.~Shchemelinin, D.~Vartsky and A.~Breskin,
  \emph{{First demonstration of VUV-photon detection in liquid xenon with THGEM
  and GEM-based Liquid Hole Multipliers}},
  \href{https://doi.org/10.1016/j.nima.2016.05.105}{\emph{Nucl. Instrum. Meth.
  A} {\bfseries 845} (2017) 218}
  [\href{https://arxiv.org/abs/1603.07669}{{\ttfamily 1603.07669}}].

\bibitem{Erdal:2017yiu}
E.~Erdal, L.~Arazi, A.~Tesi, A.~Roy, S.~Shchemelinin, D.~Vartsky et~al.,
  \emph{{Recent Advances in Bubble-Assisted Liquid Hole Multipliers in Liquid
  Xenon}}, \href{https://doi.org/10.1088/1748-0221/13/12/P12008}{\emph{JINST}
  {\bfseries 13} (2018) P12008}
  [\href{https://arxiv.org/abs/1708.06645}{{\ttfamily 1708.06645}}].

\bibitem{Erdal:2018bjg}
E.~Erdal, A.~Tesi, D.~Vartsky, S.~Bressler, L.~Arazi and A.~Breskin,
  \emph{{First Imaging Results of a Bubble-assisted Liquid Hole Multiplier with
  SiPM readout in Liquid Xenon}},
  \href{https://doi.org/10.1088/1748-0221/14/01/P01028}{\emph{JINST} {\bfseries
  14} (2019) P01028} [\href{https://arxiv.org/abs/1812.00780}{{\ttfamily
  1812.00780}}].

\bibitem{Erdal:2019dkk}
E.~Erdal, L.~Arazi, A.~Breskin, S.~Shchemelinin, A.~Roy, A.~Tesi et~al.,
  \emph{{Bubble-assisted Liquid Hole Multipliers in LXe and LAr: towards
  \textquotedblleft{}local dual-phase TPCs\textquotedblright{}}},
  \href{https://doi.org/10.1088/1748-0221/15/04/C04002}{\emph{JINST} {\bfseries
  15} (2020) C04002} [\href{https://arxiv.org/abs/1912.10698}{{\ttfamily
  1912.10698}}].

\bibitem{Erdal:2019wfb}
E.~Erdal, A.~Tesi, A.~Breskin, D.~Vartsky and S.~Bressler, \emph{{First
  demonstration of a bubble-assisted Liquid Hole Multiplier operation in liquid
  argon}}, \href{https://doi.org/10.1088/1748-0221/14/11/P11021}{\emph{JINST}
  {\bfseries 14} (2019) P11021}
  [\href{https://arxiv.org/abs/1908.04974}{{\ttfamily 1908.04974}}].

\bibitem{Breskin_2022}
A.~Breskin, \emph{Novel electron and photon recording concepts in noble-liquid
  detectors},
  \href{https://doi.org/10.1088/1748-0221/17/08/P08002}{\emph{Journal of
  Instrumentation} {\bfseries 17} (2022) P08002}.

\bibitem{Chepel_2023}
V.~Chepel, G.~Martinez-Lema, A.~Roy and A.~Breskin, \emph{First results on fhm
  — a floating hole multiplier},
  \href{https://doi.org/10.1088/1748-0221/18/05/P05013}{\emph{Journal of
  Instrumentation} {\bfseries 18} (2023) P05013}.

\bibitem{lidine_2022}
{G. Mart\'inez-Lema, A. Roy, A. Breskin and L.Arazi}, ``Progress with the
  bubble-free liquid hole-multiplier for dual-phase scintillation- \&
  electroluminescence-photon detectors.'' , presented at LIDINE 2022
  \url{https://events.camk.edu.pl/event/47/contributions/377/}.

\bibitem{mpgd_2022}
{A. Roy, G. Mart\'inez-Lema, A. Breskin and L.Arazi}, ``The mpgd-based
  bubble-free liquid hole-multiplier concept for charge and light detection in
  dual-phase noble-liquid tpcs.'' , presented at MPGD 2022
  \url{https://indico.cern.ch/event/1219224/contributions/5130450/}.

\bibitem{BONDAR2019162431}
A.~Bondar, A.~Buzulutskov, E.~Frolov, V.~Oleynikov, E.~Shemyakina and
  A.~Sokolov, \emph{Electron transport and electric field simulations in
  two-phase detectors with thgem electrodes},
  \href{https://doi.org/https://doi.org/10.1016/j.nima.2019.162431}{\emph{Nuclear
  Instruments and Methods in Physics Research Section A: Accelerators,
  Spectrometers, Detectors and Associated Equipment} {\bfseries 943} (2019)
  162431}.

\bibitem{BUZULUTSKOV201829}
A.~Buzulutskov, E.~Shemyakina, A.~Bondar, A.~Dolgov, E.~Frolov, V.~Nosov
  et~al., \emph{Revealing neutral bremsstrahlung in two-phase argon
  electroluminescence},
  \href{https://doi.org/https://doi.org/10.1016/j.astropartphys.2018.06.005}{\emph{Astroparticle
  Physics} {\bfseries 103} (2018) 29}.

\bibitem{DarkSide:2020oas}
{\scshape DarkSide} collaboration, \emph{{SiPM-matrix readout of two-phase
  argon detectors using electroluminescence in the visible and near infrared
  range}}, \href{https://doi.org/10.1140/epjc/s10052-020-08801-2}{\emph{Eur.
  Phys. J. C} {\bfseries 81} (2021) 153}
  [\href{https://arxiv.org/abs/2004.02024}{{\ttfamily 2004.02024}}].

\bibitem{BONDAR2013213}
A.~Bondar, A.~Buzulutskov, A.~Dolgov, A.~Grebenuk, E.~Shemyakina, A.~Sokolov
  et~al., \emph{First demonstration of thgem/gapd-matrix optical readout in a
  two-phase cryogenic avalanche detector in ar},
  \href{https://doi.org/https://doi.org/10.1016/j.nima.2013.07.089}{\emph{Nuclear
  Instruments and Methods in Physics Research Section A: Accelerators,
  Spectrometers, Detectors and Associated Equipment} {\bfseries 732} (2013)
  213}.

\bibitem{Doke_2002}
T.~Doke, A.~Hitachi, J.~Kikuchi, K.~Masuda, H.~Okada and E.~Shibamura,
  \emph{Absolute scintillation yields in liquid argon and xenon for various
  particles}, \href{https://doi.org/10.1143/JJAP.41.1538}{\emph{Japanese
  Journal of Applied Physics} {\bfseries 41} (2002) 1538}.

\bibitem{10.1063/5.0059497}
Y.~Zhou and P.~Zhang, \emph{{Quantum efficiency of photoemission from biased
  metal surfaces with laser wavelengths from UV to NIR}},
  \href{https://doi.org/10.1063/5.0059497}{\emph{Journal of Applied Physics}
  {\bfseries 130} (2021) 064902}
  [\href{https://arxiv.org/abs/https://pubs.aip.org/aip/jap/article-pdf/doi/10.1063/5.0059497/15266190/064902\_1\_online.pdf}{{\ttfamily
  https://pubs.aip.org/aip/jap/article-pdf/doi/10.1063/5.0059497/15266190/064902\_1\_online.pdf}}].

\bibitem{APRILE1994328}
E.~Aprile, A.~Bolotnikov, D.~Chen, R.~Mukherjee, F.~Xu, D.~Anderson et~al.,
  \emph{Performance of csi photocathodes in liquid xe, kr, and ar},
  \href{https://doi.org/https://doi.org/10.1016/0168-9002(94)91316-1}{\emph{Nuclear
  Instruments and Methods in Physics Research Section A: Accelerators,
  Spectrometers, Detectors and Associated Equipment} {\bfseries 338} (1994)
  328}.

\bibitem{garfield}
H.~Schindler, ``Garfield++.'' \url{https://gitlab.cern.ch/garfield/garfieldpp}.

\bibitem{Szydagis_2011}
M.~Szydagis, N.~Barry, K.~Kazkaz, J.~Mock, D.~Stolp, M.~Sweany et~al.,
  \emph{Nest: a comprehensive model for scintillation yield in liquid xenon},
  \href{https://doi.org/10.1088/1748-0221/6/10/P10002}{\emph{Journal of
  Instrumentation} {\bfseries 6} (2011) P10002}.

\bibitem{szydagis_m_2023_7577399}
M.~Szydagis, E.~Brown, N.~Carrara, A.~Kamaha, E.~Kozlova, D.~McKinsey et~al.,
  \emph{Noble element simulation technique},  Jan., 2023.
\newblock 10.5281/zenodo.7577399.

\bibitem{Pitt_2018}
M.~Pitt, P.~Correia, S.~Bressler, A.~Coimbra, D.S.~Renous, C.~Azevedo et~al.,
  \emph{Measurements of charging-up processes in thgem-based particle
  detectors},
  \href{https://doi.org/10.1088/1748-0221/13/03/P03009}{\emph{Journal of
  Instrumentation} {\bfseries 13} (2018) P03009}.

\bibitem{XENON:2017lvq}
{\scshape XENON} collaboration, \emph{{The XENON1T Dark Matter Experiment}},
  \href{https://doi.org/10.1140/epjc/s10052-017-5326-3}{\emph{Eur. Phys. J. C}
  {\bfseries 77} (2017) 881}
  [\href{https://arxiv.org/abs/1708.07051}{{\ttfamily 1708.07051}}].

\end{thebibliography}\endgroup

\end{document}